\documentclass[12pt, russian]{article}
\usepackage[cp1251]{inputenc}
\usepackage[T2A]{fontenc}
\usepackage{amsmath}
\usepackage{babel}
\usepackage{epsfig}
\usepackage{graphicx}

\newcommand{\be}[1]{\begin{equation}{#1}\end{equation}}

\author{ A.A. Meldianov \v{c} }

\date{ \today }
\begin{document}
\begin{center}

\section*{THE q-DEFORMED HARMONIC OSCILLATOR, COHERENT STATES,
AND THE UNCERTAINTY RELATION}
\begin{center}
\end{center}
V.V Eremin , A.A. Meldianov \footnote{ E-mail: meld@mail.ru}\\
\begin{center}
\end{center}
Lomonosov Moscow State University,Moscow, Russia\\

\end{center}

For a $q$-deformed harmonic oscillator, we find explicit coordinate
representations of the creation and annihilation operators,
eigenfunctions, and coherent states (the last being defined as
eigenstates of the annihilation operator). We calculate the product
of the “coordinate–momentum” uncertainties in qoscillator
eigenstates and in coherent states. For the oscillator, this product
is minimum in the ground state and equals 1/2, as in the standard
quantum mechanics. For coherent states, the $q$-deformation results
in a violation of the standard uncertainty relation; the product of
the coordinate- and momentumoperator uncertainties is always less
than 1/2. States with the minimum uncertainty, which tends to zero,
correspond to the values of $\lambda$ near the convergence radius of
the $q$-exponential.

\textbf{Keywords:} $q$-deformation, harmonic oscillator, creation
and annihilation operators, coherent states, uncertainty relation

\section{Introduction}
In the last decade, quantum groups and $q$-deformed algebras have
been the subject of intense investigation. After the first works
where the notion of a $q$-deformed harmonic oscillator was
introduced \cite{c1,c2,c3}, quantum groups and $q$-deformations have
found applications in various branches of physics and chemistry. In
particular, they were used to describe electronic conductance in
disordered metals and doped semiconductors \cite{c4}, to analyze the
phonon spectrum in $^4$He \cite{c5}, and to characterize the
oscillatory-rotational spectra of diatomic \cite{c6} and multiatomic
molecules \cite{c7}. As a natural application of quantum groups,
$q$-deformed quantum mechanics \cite{c8,c9} was developed, generalizing
the standard quantum mechanics based on the Heisenberg commutation
relation (the Heisenberg algebra). Quantum $q$-analogues of several
fundamental notions and models in quantum mechanics.phase space
\cite{c10}, uncertainty relation \cite{c11,c12}, density matrix \cite{c13}, harmonic
oscillator \cite{c1,c2,c10}, hydrogen atom \cite{c14}, creation and
annihilation operators, coherent states \cite{c1,c2,c15}, and dynamics
\cite{c16}.have been constructed such that they reduce to their standard
counterparts as $q\to1$. It turned out that the properties of some of
the generalized $q$-variables are notably different from the
properties of the standard quantum mechanical analogues. For
example, the $q$-oscillator has a strongly nonlinear spectrum, and
the product of uncertainties can be less than 1/2 in some cases
\cite{c11}. The properties of quantum groups (the $q$-analogues of
standard groups used in quantum mechanics) have been studied at the
abstract level in sufficient detail, and a large number of operator
identities have been obtained based on the corresponding commutation
relations. On the other hand, coordinate realizations of operators
and wave functions have received much less attention: to the best of
our knowledge, only the pioneering paper \cite{c1} contains indications
about the explicit form of the $q$-oscillator wave functions. In this paper, we obtain explicit coordinate representations for the creation and annihilation operators, wave functions, and coherent states of the $q$-deformed harmonic oscillator and investigate their dependence on the deformation parameter $q$. For coherent states, we calculate the product of the 'coordinate-momentum' uncertainties explicitly and show that it can take values arbitrarily close to zero.
\section{The $q$"~deformed harmonic oscillator}

\qquad The q-deformed harmonic oscillator (simply
q-oscillator in what follows) is described by the Hamiltonian
\begin{equation}\label{h}
    H=\frac{\hbar\omega}{2}(aa^{+}+a^{+}a)
\end{equation}
where $\omega$ - is the oscillator frequency and, $a^+$ and $a$ - are creation and annihilation operators satisfying the
commutation relation
\begin{equation}\label{c}
   [a,a^{+}]_{q}=aa^{+}-qa^{+}a=1
\end{equation}
with the deformation parameter $q$ taking values in $(0, 1)$. In what follows, we use the system of units where $\hbar=1$. In the space of eigenvectors $|n\rangle,\;n=0, 1, 2,...$ of Hamiltonian (\ref{h}) \be{H|n\rangle=E_n|n\rangle.}
the operators $a$ and $a^{+}$ act as ladder operators
\be{a|n\rangle=\sqrt{[n]}|n-1\rangle \label{an}}
\be{a^{+}|n\rangle=\sqrt{[n+1]}|n+1\rangle \label{an1}} where
$[n]=\frac{1-q^n}{1-q}$ - is the $q$"~integer \cite{c16}. The spectrum of Hamiltonian (\ref{h}) has the form
\begin{equation}\label{enn}
    E_n=\omega\left([n]+\frac{q^n}{2}\right)=\omega\left(\frac{[n]+[n+1]}{2}\right)
\end{equation}
and is bounded from above by the value
\begin{equation}\label{einfty}
    E_\infty=\frac{\omega}{1-q}
\end{equation}
At $q=1$ it becomes the spectrum of the standard harmonic oscillator
\begin{equation}\label{eob}
    E_n=\omega\left(n+\frac{1}{2}\right)
\end{equation}
For small deviations from unity $(q = 1 - \varepsilon)$, the spectrum becomes quadratic,
\begin{equation}\label{eps}
    E_n=\omega\left(n+\frac{1}{2}-\frac{n^2}{2}\varepsilon+O(\varepsilon^2)\right)
\end{equation}
which allows using the $q$-oscillator to describe oscillatory-rotational levels of diatomic \cite{c6} and multiatomic \cite{c7} molecules with the unharmonicity parameter $\varepsilon$.
As for the standard harmonic oscillator, the eigenvectors of
Hamiltonian (\ref{h}) can be found by consecutively applying the creation
operator $a^+$ to the vacuum state vector:
\begin{equation}\label{ketn}
    |n\rangle=\frac{1}{\sqrt{[n]}}a^+|n-1\rangle=\ldots=\frac{1}{\sqrt{[n]!}}(a^+)^n|0\rangle
\end{equation}
The vectors thus constructed are orthonormalized and constitute a
complete system. Energy levels (\ref{enn}) are determined only by the form
of Hamiltonian (\ref{h}) and by commutation relation (\ref{c}). To find the
eigenvectors explicitly, we must specify the coordinate representation for the creation and annihilation operators. We suggest taking it in the form
\be{a=\frac{\exp({-2{i\alpha}x})-\exp({{i\alpha}\frac{d}{dx}})\exp({-{i\alpha}
x})}{-i\sqrt{1-\exp({-2\alpha^2})}}\label{qa}}
\be{a^{+}=\frac{\exp({2{i\alpha}x})-\exp({{i\alpha}x})\exp({{i\alpha}
\frac{d}{dx}})}{i\sqrt{1-\exp({-2\alpha^2})}}\label{qak}} где $\alpha
=\sqrt{-\ln{q}/2}$, $0<\alpha<\infty$.
This representation satisfies
relation (\ref{c}) and yields the standard harmonic-oscillator expressions
in the limit as $q \to 1$ ($\alpha\to 0$). The corresponding coordinate and
momentum operators in q-deformed quantum mechanics are given by
\begin{eqnarray}
    \hat{x}&=&\frac{a+a^+}{\sqrt{2}}=\sqrt{\frac{2}{1-\exp(-2\alpha^2)}}\times\nonumber\\ &\times&{\left(\sin{(2\alpha
x)}-\exp\left({\frac{\alpha^2}{2}}\right)\sin{\left(\alpha
x+\frac{\alpha^2}{2}i\right)}\exp\left({{i\alpha} \frac{d}{dx}}\right)\right)}\nonumber \\
\hat{p}&=&\frac{a-a^+}{i\sqrt{2}}=\sqrt{\frac{2}{1-\exp(-2\alpha^2)}}\times\nonumber\\
&\times&{\left(\cos{(2\alpha x)}-\exp\left({\frac{\alpha^2}{2}}\right)\cos{\left(\alpha
x +\frac{\alpha^2}{2}i\right)}\exp\left({{i\alpha} \frac{d}{dx}}\right)\right)}
\end{eqnarray}\label{qxp}

Using the proposed coordinate representation, we first find the vacuum-state wave function. The normalized solution of the equation $a|0\rangle=0$, i.e.,
\begin{equation}\label{vac}
    \left[\exp({-2{i\alpha}x})-\exp\left({{i\alpha}\frac{d}{dx}}\right)\exp({-{i\alpha}
x})\right]\Psi_0(x)=0
\end{equation}
is given by
\begin{equation}\label{psi0}
    \Psi_0(x)=\frac{1}{\pi^{\frac{1}{4}}}\exp\left(-\frac{x^2}{2}+\frac{3}{2}i\alpha
    x\right)
\end{equation}
The probability density of the vacuum state $|\Psi_0(x)|^2$ has a Gaussian shape and is independent of the deformation parameter $q$.

Substituting relations (\ref{qak}) and (\ref{psi0}) in formula (\ref{ketn}),
we obtain an explicit coordinate expression for the normalized
$q$"~oscillator wave functions:
\begin{eqnarray}\label{psin}
    |n\rangle=\Psi_n(x)=\frac{\exp({-\frac{x^2}{2}+\frac{3}{2}{i\alpha
}x})}{\pi^{\frac{1}{4}}i^n{(1-\exp({-2\alpha^2}))}^{\frac{n}{2}}\sqrt{[n]!}
}\times\nonumber\\ \times\sum_{k=0}^n\frac{{(-1)}^k[n]!}{[k]![n-k]!}\exp({(n-k)2{i\alpha
}x-k\alpha^2} )
\end{eqnarray}
where $[n]!=[1][2]...[n]$. As $q\to1$ $(\alpha\to0)$, wave functions (\ref{psin})converge to wave functions of the harmonic oscillator,
\begin{equation}\label{ps}
    \Psi_n(x)\stackrel{q\to1}{\longrightarrow}\frac{1}{\pi^{\frac{1}{4}}2^{\frac{n}{2}}\sqrt{n!}}H_n(x)\exp\left(-\frac{x^2}{2}\right)
\end{equation}
and as $q\to0$ $(\alpha\to\infty)$ hey ``degenerate'' in that the probability density for any $n$ tends to the ground-state probability density,
\begin{equation}\label{p0}
    \Psi_n(x)\stackrel{q\to0}{\longrightarrow}\frac{1}{\pi^{\frac{1}{4}}i^n}\exp\left(-\frac{x^2}{2}+\left(2n+\frac{3}{2}\right)i\alpha x\right)
\end{equation}
\begin{equation}\label{absp0}
    |\Psi_n(x)|^2\stackrel{q\to0}{\longrightarrow}\frac{1}{\pi^{\frac{1}{2}}}\exp\left(-x^2\right)
\end{equation}

It is interesting to estimate the ``coordinate-momentum''
uncertainty relation for the $q$"~oscillator eigenstates. Using ladder
relations (\ref{an}) and (\ref{an1}), we find the means of the coordinate and
momentum operators:
\begin{eqnarray}\label{sx}
    \langle x\rangle=\langle p\rangle=0
\nonumber\\
    \langle x^2\rangle=\langle p^2\rangle=\frac{E_n}{\omega}
\end{eqnarray}
This implies that the product of the coordinate and momentum
uncertainties for the pure $n$th state of the $q$"~oscillator is equal to
the energy of this state in the units of $\omega$:
\begin{equation}\label{delta}
    \Delta x\Delta p=\left(\langle x^2\rangle -\langle x\rangle^2\right)^{\frac{1}{2}}\left(\langle p^2\rangle -\langle
p\rangle^2\right)^{\frac{1}{2}}=\frac{E_n}{\omega}\leq n+\frac{1}{2}
\end{equation}
It is less than the
analogous value for the standard harmonic oscillator for all $n$ except $n = 0$. At $n = 0$ , the product of the uncertainties is minimum
and is given by 1/2.

The idea that deformed commutation relations
can lead to a change in the uncertainty relation was first proposed
back in \cite{c20}. The first estimates showing that the product of the
coordinate and momentum uncertainties for a $q$-analogue of the
Heisenberg algebra can be less than 1/2 were given in \cite{c11,c13}. To
find the minimum value of the product of uncertainties, we turn to
the $q$"~analogue of coherent states of a harmonic oscillator.

\section{Coherent states}
Coherent states In quantum mechanics, the minimum possible product
of uncertainties is characteristic of the coherent states, one of
whose definitions involves the annihilation operator for the
harmonic oscillator \cite{c18}. We consider a generalization of this
notion to the case of $q$"~deformations and calculate the coordinate
and momentum uncertainties in a coherent state.
We define a coherent state as an eigenstate of the annihilation operator,
\begin{equation}\label{l}
    a|\lambda\rangle=\lambda|\lambda\rangle
\end{equation}
where $\lambda$ - is a complex number subjected to some restrictions
below. With expression (\ref{an}), we can show that this definition is
equivalent to the decomposition
\begin{equation}\label{lr}
    |\lambda\rangle=c_0\sum_{n=0}^{\infty}\frac{\lambda^n}{\sqrt{[n]!}}|n\rangle
\end{equation}
in the $q$"~oscillator eigenfunctions or to the action of the shift
operator on the vacuum state,
\begin{equation}\label{lra}
    |\lambda\rangle=c_0\sum_{n=0}^{\infty}\frac{{(\lambda
a^+)}^n}{[n]!}|0\rangle=c_0\exp_q\left(\lambda a^+\right)|0\rangle
\end{equation}
where $\exp_q$ is the $q$"~exponential \cite{c23}. The constant $c_0$ is determined from the normalization condition $\langle\lambda|\lambda\rangle=1$:
\begin{equation}\label{c0}
    |c_0|^2=\frac{1}{\sum\limits_{n=0}^{\infty}\frac{{|\lambda
|^{2n}}}{[n]!}}=\frac{1}{\exp_q(|\lambda |^2)}
\end{equation}

The existence of a convergence disk for the $q$"~exponential imposes a restriction on the coherent states,
\begin{equation}\label{r}
    |\lambda|<\frac{1}{\sqrt{1-q}}
\end{equation}
in contrast to the case of the standard quantum mechanics, where $\lambda$
can take any values. To calculate the means of the coordinate and
momentum, we need the energy of the coherent state:
\begin{equation}\label{vkk}
 \langle\lambda |H|\lambda\rangle= \langle\lambda |\frac{1}{2}+\frac{1+q}{2}a^+a|\lambda\rangle=\frac{1}{2}+\frac{1+q}{2}|\lambda|^2
\end{equation}

Relations (\ref{vkk}) and (\ref{r}) imply that the energy of $q$-coherent states is bounded:
\begin{equation}
    \langle\lambda |H|\lambda\rangle <\frac{1}{1-q}
\end{equation}

Calculating the other means proceeds from coherent-state definition (\ref{l}):

\begin{eqnarray}\label{arr}
\langle x\rangle &=&\langle
    \lambda |\frac{a+a^+}{\sqrt{2}}|\lambda\rangle=\frac{\lambda+\lambda^*}{\sqrt{2}}\nonumber\\
\langle p\rangle &=&\langle
    \lambda
|\frac{a-a^+}{i\sqrt{2}}|\lambda\rangle=\frac{\lambda-\lambda^*}{i\sqrt{2}}\nonumber\\
\langle x^2\rangle &=&\langle \lambda
|\left(\frac{a+a^+}{\sqrt{2}}\right)^2|\lambda\rangle=\frac{1}{2}\left(\lambda^2+{\lambda^{*}}^2\right)+\langle\lambda|H|\lambda\rangle\nonumber\\
\langle p^2\rangle &=&\langle \lambda
|-\left(\frac{a-a^+}{\sqrt{2}}\right)^2|\lambda\rangle=-\frac{1}{2}\left(\lambda^2+{\lambda^{*}}^2\right)+\langle\lambda|H|\lambda\rangle
\end{eqnarray}
\\From (\ref{arr}) and (\ref{l}) with (\ref{vkk}) taken into account, we find the product of the coordinate and momentum uncertainties for the coherent state,  $|\lambda\rangle$:
\begin{equation}\label{deltalambda}
    \Delta x\Delta
p=\langle\lambda|H|\lambda\rangle-|\lambda|^2=\frac{1}{2}-\frac{1-q}{2}|\lambda|^2
\end{equation}
For any $q < 1$ and admissible $\lambda$ , this product is less than the standard quantum mechanical value 1/2. Moreover, it can be made arbitrarily small as
$\lambda$ increases. Therefore, while the coherent states in the standard
quantum mechanics are defined as the minimum-uncertainty states,
this definition does not work in the $q$"~deformed quantum mechanics:
the minimum value $\Delta x\Delta p$, equal to zero in the $q$"~deformed case (see
(\ref{deltalambda})), is unreachable because the $q$"~exponential series does not
converge at the corresponding values of $\lambda$
$(|\lambda|={(1-q)}^\frac{1}{2})$

To find the wave functions of coherent states explicitly, we solve Eq.
(\ref{l}), which involves annihilation operator (\ref{qa}):
\begin{equation}\label{ur}
    \left(\frac{\exp({-2{i\alpha}x})-\exp({{i\alpha}\frac{d}{dx}})\exp({-{i\alpha}
x})}{-i\sqrt{1-\exp({-2\alpha^2})}}-\lambda\right)\Psi_{\lambda}(x)=0
\end{equation}
We seek the solution in the form
\begin{equation}\label{pods}
    \Psi_{\lambda}(x)=f(\exp(2i\alpha x))\Psi_{0}(x)
\end{equation}
where the zeroth coherent state has form (\ref{psi0}). Substituting expression (\ref{pods}) in (\ref{ur})and performing some transformations, we obtain
\begin{equation}\label{ur1}
    \frac{f(A)-f(qA)}{A(1-q)}=\frac{\lambda}{i\sqrt{1-q}}f(A)
\end{equation}
where $A\equiv \exp(2i\alpha x)$. Equation (\ref{ur1}) is a
$q$"~analogue of the standard differential equation for the exponential
\cite{c23} and has the solution
\begin{equation}\label{ff1}
    f(A)=\exp_q\left(\frac{\lambda}{i\sqrt{1-q}}A\right)
\end{equation}
Substituting this solution in  (\ref{ff1}) в (\ref{pods})and recalling the normalization conditions, we find the explicit form of coherent states:
\begin{eqnarray}\label{coh}
    \Psi_{\lambda}(x)=\frac{1}{\pi^{\frac{1}{4}}{\left[\exp_q(|\lambda|^2)\right]}^{\frac{1}{2}}}\exp_q\left(i\frac{\lambda\sqrt{q}}{\sqrt{1-q}}\right)\times\nonumber\\ \times \exp_q\left(\frac{\lambda}{i\sqrt{1-q}}\exp(2i\alpha x)\right)\exp\left(-\frac{x^2}{2}+\frac{3}{2}i\alpha
    x\right)
\end{eqnarray}
As $q\to1$ , this expression reduces to the coherent states of the standard quantum oscillator:
\begin{equation}\label{limcoh}
    \Psi_{\lambda}(x)\stackrel{q\to1}{\longrightarrow}\frac{1}{\pi^{\frac{1}{4}}}\exp\left(-\frac{{(x-\lambda\sqrt{2})}^2}{2}\right)
\end{equation}
It can be shown that expression (\ref{coh}) is
equivalent to decomposition (\ref{lr}) with respect to the $q$"~oscillator
eigenfunctions. The probability density of coherent states is given
by deformed Gaussian curves whose amplitude increases as $q$
decreases, i.e., as the deformation increases. The structure of the
curves is most interesting in the case of large deformations $(q << 1)$
and large $\lambda$, i.e., for small-uncertainty states (\ref{deltalambda}). As $\lambda$ increases from zero to the limit value $\frac{1}{\sqrt{1-q}}$, the original Gaussian curve (\ref{psi0}) is deformed into several localized narrow peaks.
To explain this, we consider the limit behavior of coherent states
as $q\to0$. Formula (\ref{coh}) and the properties of the $q$"~exponential \cite{c23} imply that
\begin{equation}\label{0}
    |\Psi_{\lambda}(x)|^2\stackrel{q\to0}{\longrightarrow}\frac{1}{\pi^{\frac{1}{2}}}\frac{1-|\lambda|^2}{1+|\lambda|^2-2{\rm Im}(\lambda \exp(2i\alpha
x))}\exp(-x^2)
\end{equation}
As $q\to0$ $(\alpha\to\infty)$ , the rapidly oscillating term in the denominator cuts separate peaks from the Gaussian function.

\section{Conclusions}
The main results in this work are the explicit coordinate representations for the $q$"~deformed oscillator annihilation and creation operators (Eqs. (\ref{qa}) and (\ref{qak})), eigenfunctions (Eq. (\ref{psin})), and coherent states (Eq. (\ref{coh})). Another important result is that the $q$"~oscillator coherent states minimize the ``coordinate-momentum'' uncertainty relation in Eq. (\ref{deltalambda}), which, in contrast to the standard quantum mechanics, is not bounded by the value 1/2 but can take arbitrarily small positive values
depending on the deformation parameter $q$ and the coherent-state
eigenvalue $\lambda$. States with the minimum uncertainty, tending to zero,
correspond to the values of $\lambda$ near the convergence radius of the
$q$"~exponential: $|\lambda|\to\frac{1}{\sqrt{1-q}}$.

To conclude, we discuss the physical interpretation of the deformation parameter $q$, with possible applications of our results in mind. The structure of quantum groups is naturally related to some exactly solvable models in statistical mechanics and field theory \cite{c1}, but the meaning of deformation of the standard quantum mechanical relations is still unclear. It has been conjectured that this deformation may be observed at Planckian distances (shorter than $10^{-17}$ cm \cite{c2,c11}), but these conjectures cannot yet be verified experimentally. On the other hand, the deformation parameter can be given a less
fundamental interpretation in molecular physics and quantum chemistry, based on nonlinear properties of the $q$"~oscillator.
Man'ko and coauthors \cite{c24} already noted that the $q$"~oscillator can
be regarded as the standard nonlinear oscillator with a peculiar
dependence of the frequency on the oscillation amplitude. Later, it
was established that quantum dynamics of the $q$"~oscillator is
isomorphic to the dynamics of the standard nonlinear oscillator with
quadratic unharmonicity \cite{c19}. Therefore, it appears quite reasonable
to interpret the q parameter as an unharmonicity measure of
molecular oscillations. This is further supported by the
nonlinearity of the $q$"~oscillator spectrum (see (\ref{eps})), which at small
deformations is quadratic, as are oscillatory spectra of electronic
states of many molecules. We therefore hope that the results in this
work, in particular, the explicit form of the $q$"~oscillator wave
functions, can be used in the theory of molecular spectra, for
example, to calculate the line intensities determined by the overlap
integrals of wave functions.

Acknowledgments. This paper was supported by the Russian Foundation
for Basic Research (Grant No. 03-03-32521).

 \end{document}